\documentclass{elsart}
\usepackage{natbib}
\usepackage{epsfig}

\begin{document}
\runauthor{Zhou, Xu, Wu, Wang, Hong}
\begin{frontmatter}

\title{Quakes in Solid Quark Stars}

\author[CAS-Shanghai]{A. Z. Zhou},
\author[Beijing]{R. X. Xu}\footnote{
Corresponding author.\\
{\em Email address:} rxxu@bac.pku.edu.cn (R. X. Xu).},
\author[Beijing]{X. J. Wu},
\author[CAS-U]{N. Wang},
\author[CAS-Shanghai]{X. Y. Hong}

\address[CAS-Shanghai]{Shanghai Astronomical Observatory,
Chinese Academy of Sciences, Shanghai 200030, China}

\address[Beijing]{School of Physics, Peking University,
Beijing 100871, China}

\address[CAS-U]{Urumqi Station, the National Astronomical
Observatories, the Chinese Academy of Sciences, Urumqi 830011}

\begin{abstract}

A starquake mechanism for pulsar glitches is developed in the
solid quark star model. It is found that the general glitch
natures (i.e., the glitch amplitudes and the time intervals) could
be reproduced if solid quark matter, with high baryon density but
low temperature, has properties of shear modulus $\mu=10^{30\sim
34}$ erg/cm$^3$ and critical stress $\sigma_{\rm c}=10^{18\sim
24}$ erg/cm$^3$. The post-glitch behavior may represent a kind of
damped oscillations.

\vspace{5mm} \noindent {\it PACS codes:} 97.60.G, 97.60.J, 11.80.F

\end{abstract}

\begin{keyword}
pulsars, neutron stars, elementary particles
\end{keyword}

\end{frontmatter}


\section{Introduction}

Pulsars are unique objects, with which all types of elemental
interaction could be tested extremely.
However, the most elementary question relevant is still open: {\em
What is the nature of pulsars?} It is conventionally thought that
pulsars are simply a kind of boring big ``nuclei'' --- neutron
stars, but more and more attention is paid to the quark star model
for pulsars \citep{xzq02,xu03a} since {\em no} convincing work,
neither in theory from first principles nor in observation, has
confirmed Baade-Zwicky's original idea that supernovae produce
neutron stars.
The bare quark surface is suggested to be a new probe for
identifying quark stars with strangeness, and possible
observational evidence for bare strange stars appears: the
drifting subpulses of radio pulsars, ultra-high luminosity of soft
$\gamma$-ray repeaters, non-atomic thermal spectra of isolated
``neutron'' stars \citep{xu03b}.
However, can the bare strange star model reproduce most of the
general features of pulsars (especially glitches)?

The observation of free precession in PSR B1828-11 \cite{sls00}
and PSR B1642-03 \citep{slu01} challenges astrophysicists today to
re-consider the internal structure of radio pulsars
\citep{horv04}.
The current model for glitches involves neutron superfluid vertex
pinning and the consequent fluid dynamics.
However, the pinning should be much weaker than predicted in the
glitch models, otherwise the vortex pinning will damp out the
precession on timescales being much smaller than observed. In
addition the picture, that a neutron star core containing
coexisting neutron vertices and proton flux tubes, is also
inconsistent with observations of freely precessing pulsars
\citep{link03}.
It is then supposed that the hydrodynamic forces presented in a
precessing star are probably sufficient to unpin all of the
vortices of the inner crust \citep{lc02} since a definitive
conclusion on the nature of vertex pinning has not been reached
yet due to various uncertainties in the microscopic physics.
But recently, Levin \& D'Angelo \citep{ld04} studied the
magnetohydrodynamic (MHD) coupling between the crust and the core
of a rotating neutron star, and found that the precession of PSR
B1828-11 should decay over a human lifetime. This well-defined MHD
dissipation should certainly be important in order to test the
stellar models.

An alternative way to understand both glitch and free-precession
could be through the suggestion that radio pulsars are solid quark
stars \citep{xu03,xu04}.
A solid quark star is just a rigid-like body, no damping occurs,
and the solid pulsar model may survive future observational tests
if the free precession keeps the same over several tens of years.
A neutron star could not be in a solid state, whereas a cold quark
star could be.
Such a solid state of quark matter could be very probably
Skyrme-like\footnote{%
Skyrme \citep{skyrme62} considered baryons as solitons. The
$n-$quark clusters might also be described as solitons in a
similar way.
} \citep{ob99,lee03}, %
the study of which may help us to understand dense quark matter
with low temperature.

Fluid strange-star (even with possible crusts) models were noted
to be inconsistent with the observations of pulsar glitches more
than one decade ago \citep{alpar87}. Modifications with the
inclusion of possible stable particles to form a differentiated
structure of so-called strange pulsars was also suggested
\citep{benv90}, but is not popular because of a disbelief in the
employed physics \citep{hps93,horv04}.
However, can a fully solidified quark star proposed \citep{xu03}
really reproduce the glitch behaviors observed?
One negative issue is that giant glitches are generally not able
to occur at an observed rate in a solid neutron star \citep{bp71}.
Nonetheless, more strain energy could be stored in a solid quark
star due to an almost uniform distribution of density (the density
near the surface of a bare strange star is $\sim 4\times 10^{14}$
g/cm$^3$) and high shear modulus introduced phenomenologically for
solid quark matter with strangeness. More energy is then released,
and this may enable a solid pulsar to glitch frequently with large
amplitudes.
Furthermore, the post-glitch behaviors may represent damped
vibrations.
In this paper, we try to model glitches in a starquake scenario of
solid quark stars.

\section{The model}

A quake model for a star to be mostly solid was generally
discussed by Baym \& Pines \citep{bp71} and others
\cite{ill74,pan75}, who parameterized the dynamics for solid
crusts, and possible solid cores, of neutron stars.
Strain energy develops when a solid star spins down until a quake
occurs when stellar stresses reach a critical value.
Two possibilities were considered previously. Ruderman
\citep{rud69} assumed that the entire strain energy is relieved in
the quake, while Baym \& Pines \citep{bp71} suggested only part of
the stress is released and that the plastic flow is negligible.
We present a third possibility that, during a quake, the entire
stress is almost relieved at first when the quake cracks the star
in pieces of small size (the total released energy $E_{\rm t}$ may
be converted into thermal energy $E_{\rm therm}$ and kinematic
energy $E_{\rm k}$ of plastic flow, $E_{\rm t}=E_{\rm
therm}+E_{\rm k})$, but the part of $E_{\rm k}$ might be re-stored
by stress due to the anelastic flow (i.e., the kinetic energy is
converted to strain energy again).
As shown in Fig.1, a quark star may solidify with an initial
oblateness $\varepsilon_0$; stress (to be relative to
$\varepsilon_0$) increases as the star losses its rotation energy,
until the star reaches an oblateness $\varepsilon_{+1}$. A quake
occurs then, and the reference point of strain energy should be
changed to $\varepsilon_1$ (the oblateness of a star without shear
energy) at this moment. Damped vibrations in the potential field
(see the inserted up-right part of Fig.1) occur therefore. After a
glitch, the star solidifies and becomes elastic body again.
%
%
\begin{figure}[t]
  \centering
  \begin{minipage}[t]{.5\textwidth}
    \centering
    \includegraphics[width=9cm]{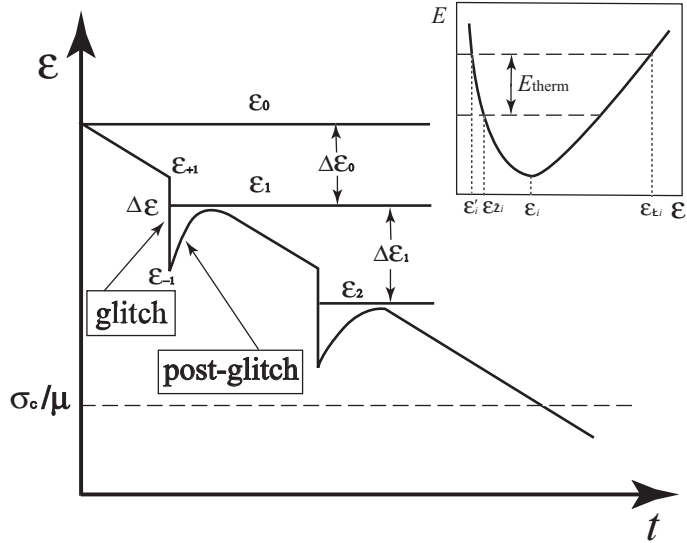}
  \end{minipage}
     \caption{
The oblateness as a function of time, $\varepsilon(t)$. The
horizontal solid lines are for the reference points of the strain.
The steps, drawn very exaggeratedly, are starquakes. No starquake
occurs anymore when $\varepsilon$ falls below $\sigma_{\rm c}/\mu$
(i.e., the stress can not be large enough for a quake).
              }
\end{figure}
No memory of the first glitch affects the second one.

The density of quark stars with mass $< \sim 1.5 M_\odot$ can be
well approximated to be uniform \citep{afo86}.
As a star, with an initial value $\varepsilon_{0}$, slows down,
the expected $\varepsilon$ decreases with increasing period.
However, the rigidity of the solid star causes it to remain more
oblate than it would be had it no resistance to shear. The strain
energy in the star reads \citep{bp71}
\begin{equation}
E_{\rm strain}= B(\varepsilon  - \varepsilon_{0} )^2,
\end{equation}
and the mean stress $\sigma$ in the star is
\begin{equation}
\sigma  = \left| {\frac{1}{{V_{} }}\frac{{\partial E_{\rm strain}
}}{{\partial \varepsilon }}} \right| = \mu (\varepsilon _0  -
\varepsilon ),%
\label{sigma}
\end{equation}
where $\varepsilon$ is stellar oblateness\footnote{
The eccentricity $e$ is defined by $e^2=1-c^2/a^2$ ($a$ and $c$
are the semimajor and semiminor axes, respectively).
The oblateness (or ellipticity) $\varepsilon\equiv (I-I_0)/I_0$
($I_0$ is the nonrotating, spherical moment of inertia) is related
to $e$ through $\varepsilon=(1-e^2)^{-1/3}-1\approx e^2/3$.
}, %
$V=4\pi R^3/3$ is the volume of the star, and $\mu  = 2B/V $ is
the mean shear modulus of the star.
We note that, in the following calculations, the stress-increase
developed by the decrease of oblateness (due to the spindown) is
only included. However, azimuthal stress due to the general
relativistic effect \citep{xu04} (maybe similar to the
frame-dragging effect in vacuum) of rotating solid stars may
contribute significantly, though, unfortunately, the theoretical
answers to elastic relativistic-stars with rotation are very
difficult to be worked out.

The total energy of such a rotating star with mass $M$ and radius
$R$ is the sum of the gravitational energy $E_{\rm gravi}$, the
rotation energy $ E_ {\rm rot}$, the strain energy $E_{\rm
strain}$, the bulk energy $E_{\rm v}$, and the surface energy
$E_{\rm s}$,
\begin{eqnarray}
E &=& E_{\rm gravi} + E_ {\rm rot} + E_{\rm strain} + E_{\rm v} +
E_{\rm s}\nonumber\\ &=& E_{\rm 0} + A\varepsilon ^2 + L^2/(2I) +
B(\varepsilon  - \varepsilon_i)^2 +
E_{\rm v} + E_{\rm s},%
\label{E}
\end{eqnarray}
where $\varepsilon_i$ is the reference oblateness before the
$(i+1)$-th glitch occurs, $E_{\rm 0}=-3M^2G/(5R)$, $I$ is the
moment of inertia, $L=I\Omega$ is the stellar angular momentum,
$\Omega=2\pi/P$ ($P$ the rotation period), and the coefficients
$A$ and $B$ measure the gravitational and strain energies
\citep{bp71}, respectively,
\begin{equation}
A = \frac{3}{{25}}\frac{{GM_{}^2 }}{R},
\end{equation}
\begin{equation}
B = {2\over 3}\pi R^3 \mu.
\end{equation}
The changes of $E_{\rm v}$ and $E_{\rm s}$ are much smaller than
that of $E_{\rm gravi}$, $E_ {\rm rot}$, or $E_{\rm strain}$ when
a star spins down, according to a mass formula for strange quark
matter to be analogous to the Bethe-Weizsacher semi-empirical mass
function \citep{dcs93}. We therefore neglect $E_{\rm v}$ and
$E_{\rm s}$ in the following calculations.

By minimizing the total energy $E$, a real state satisfies (note
that $\partial I(\varepsilon)/\partial \varepsilon=I_0$),
\begin{equation}
 \varepsilon  = \frac{{I_0 \Omega ^2}}{{4(A +
B)}} + \frac{B}{{A + B}}\varepsilon _i. %
\label{epsilon}
\end{equation}
The reference oblateness is assumed, by setting $B=0$ in
Eq.(\ref{epsilon}), to be
\begin{equation}
\varepsilon _i  = I_0 \Omega^2/(4A).%
\label{epsiloni}
\end{equation}
A star with oblateness of Eq.(\ref{epsiloni}) is actually a
Maclaurin sphere. When the star spins down to $\Omega$, the stress
develops to
\begin{equation}
\sigma = \mu [\frac{A}{{A + B}}\varepsilon_i - \frac{{I_0 \Omega^2
}}{{4(A + B)}} ],%
\label{sigmamu}
\end{equation}
according to Eq.(\ref{sigma}). A glitch takes place if the stress
is greater than a critical one, $\sigma>\sigma_{\rm c}$.

The first quake is characterized by a sudden shift of
$\varepsilon$, the amount of which is
\begin{equation}
\Delta \varepsilon =\varepsilon_{+1}-\varepsilon_{-1},
\end{equation}
where the reference point is changed from $\varepsilon_0$ to
$\varepsilon_1$.
Due to the conservation of stellar angular momentum, the sudden
change in the oblateness results in an increase of spin frequency,
\begin{equation}
\frac{{\Delta \Omega }}{\Omega } =  - \frac{{\Delta I}}{I} =
\Delta \varepsilon.%
\label{DeltaO/O}
\end{equation}
Starquakes would also produce a thermal energy dissipation during
glitches that could be expected to be observable as an increase of
X-ray luminosity soon after glitch. According to the conservation
of energy, one obtains from Eq.(\ref{E}),
\begin{eqnarray}
 A\varepsilon_{+1}^2+0.5\varepsilon_{+1}I_0\Omega_1^2+B(\varepsilon_{+1}-\varepsilon_0)^2=
 A\varepsilon_{-1}^2+~~~~~~~~~~~~\nonumber\\
 0.5\varepsilon_{-1}I_0\Omega_1^2+B(\varepsilon_{-1}-\varepsilon_1)^2+E_{\rm thermal},%
 \label{epsinon-1}
 \end{eqnarray}
where $\Omega_1$, which can be obtained by $\sigma=\sigma_{\rm c}$
from Eq.(\ref{sigmamu}), is the spin frequency when the first
quake occurs, $E_{\rm thermal}$ is the released energy in a
starquake.
The observed glitch size $\Delta\Omega/\Omega$ can be calculated
from Eq.(\ref{epsilon}) to Eq.(\ref{epsinon-1}).
The calculations, based on these equations, are parameterized by
five input quantities: $M$, $\rho$, $\mu$, $\sigma_{\rm c}$ and
$E_{{\rm thermal}}$, and two of them are fixed to be
$M=1.4M_\odot$ and $\rho=4\times 10^{14}$ g/cm$^3$.

Fig. 2 shows the results for various $\mu$ and $\sigma_{\rm c}$.
\begin{figure}[t]
  \centering
  \begin{minipage}[t]{.5\textwidth}
    \centering
    \includegraphics[width=9cm]{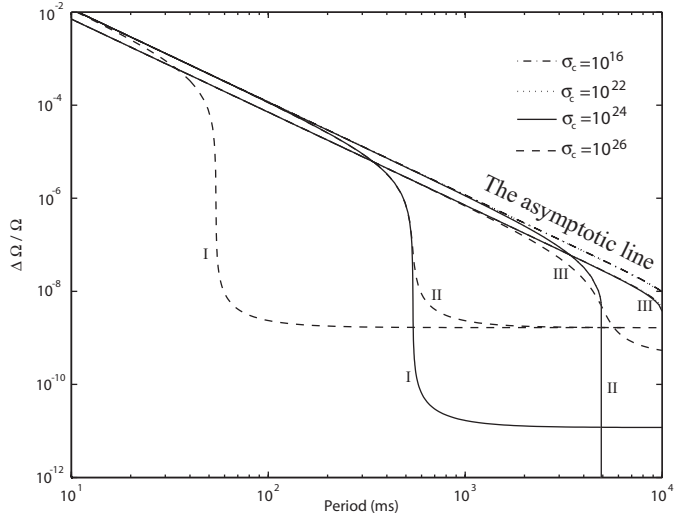}
  \end{minipage}%
     \caption{
The fractional increase in the spin frequency during a glitch,
$\Delta \Omega/\Omega$, vs. the rotation period, $P$. The lines
labelled I, II, and III are for different shear modulus, $\mu$. I:
$\mu=10^{30}$ erg/${\rm cm^{3}}$, II: $\mu=10^{32}$ erg/${\rm
cm^{3}}$, and III: $\mu=10^{34}$ erg/${\rm cm^{3}}$. The dash-dot,
dotted, solid and dashed lines are for $\sigma_{\rm c}=10^{16}$,
$\sigma_{\rm c}=10^{22}$, $\sigma_{\rm c}=10^{24}$ and
$\sigma_{\rm c}=10^{26}$ erg/${\rm cm^{3}}$, respectively. We
observe that the curves tend towards the asymptotic line when
$\sigma_{\rm c}< \sim 10^{22}$ erg/${\rm cm^{3}}$.
The stellar mass and mean density are chosen to be mass
$M=1.4M_\odot$ and density $\rho=4\times 10^{14}$ g/cm$^3$ in the
calculations.
              }
\end{figure}
We see that, for $\sigma_{\rm c}>\sim 10^{22}$ erg/${\rm cm^{3}}$,
$\Delta\Omega/\Omega$ decays to be a constant for pulsars with
large rotation periods. Small $\mu$ may generally result also in a
small jump of $\Delta\Omega/\Omega$ for a certain period $P$.
However, for $\sigma_{\rm c}<\sim 10^{22}$ erg/${\rm cm^{3}}$, the
asymptotic line is reached, which means that $\ln
(\Delta\Omega/\Omega)\propto -\ln P$.
We choose $E_{\rm therm}=10^{36}$ erg/${\rm cm^{3}}$ in
computation, but the curves do not depend sensitively on $E_{\rm
therm}$.
It is found that the glitch amplitude, $\Delta\Omega/\Omega$,
could be as high as observed if the shear modulus $\mu \sim
10^{30-34}$ erg/cm$^3$. Though the quark matter with low
temperature and high baryon density is focused recent years, it is
still impossible to determine the properties of such QCD phase by
first principles (most of the calculations are started from QCD
phenomenological models). One of the proposed states of strange
stars at low temperature could be a solid state \citep{xu03} (and
also the 2nd paragraph of \S4), but the calculations on solid
state is much more difficult than that of fluid one. Nevertheless,
we may estimate the shear modulus $\mu$ originated only by
electric interaction between charged $n-$quark clusters ($n$: the
quark number in a cluster) and a uniform background of electrons,
through a similar quantum mechanical calculation of metals
\citep{fuc36}. The effective shear modulus, averaged over
polarizations and directions, can be well fitted by $\mu\sim 0.12
N(Ze)^2/a$ in asymptotic case, where $Z$ is the charge of
quark-clusters, $N$ the cluster number density and $a$ the
separation between clusters \citep{Strohmayer91}. For strange
quark matter with baryon number density $n_{\rm B}$ and electron
number density to be $\sim 10^{-3}$ that of quarks, one comes to
\begin{equation}
\mu \simeq 10^{28} ({n_{\rm B}\over 2n_0})^{4/3} ({n\over
10^3})^{2/3}
~~{\rm erg/cm^3}, %
\label{mu-estimation}
\end{equation}
where $n_0=0.16$ baryons per fm$^3$ is the nuclear saturation
density.
This result can be regarded as a low limit modulus of solid
strange quark matter since the van der Waals-type color
interaction, with a high coupling constant (to be corresponding to
the electric charge $e$ in electromagnetic interaction), may
result in a larger shear modulus. We might then conclude that
$\mu$ in the range of $10^{30-34}$ erg/cm$^3$ is not impossible.
If the kilohertz quasi-periodic oscillations are relevant to the
global oscillation behavior of solid quark stars, the shear
modulus could be \citep{xu03} $\mu\sim 10^{32}$ erg/cm$^3$, which
is much larger than the value ($\sim 10^{28}$ erg/cm$^3$) of
neutron star crust. This could be unsurprise due to much high
density.

After the first quake, from Eq.(\ref{sigmamu}), the stress in the
star builds up again, at a rate of
\begin{equation}
\dot \sigma  =  - \mu \dot \varepsilon  =  - \frac{{\mu I_0
}}{{2(A + B)}}\Omega \dot \Omega, %
\label{sigmadot}
\end{equation}
which is almost a constant for a certain pulsar with $P$ and $\dot
P$ during a period when the effect of $\ddot{P}$ is negligible.
Another quake occurs after a time of
\begin{equation}
t_{\rm q} = \sigma_{\rm c}/{\dot \sigma}. %
\label{tq}
\end{equation}
The calculation for certain $P$ and $\dot P$ is presented in
Fig.3. For a certain set of \{$P,{\dot P}$\}, the interval $t_{\rm
q}$ become longer for a higher value of $\sigma_{\rm c}$.
It is, however, not necessary to expect that a radio pulsar should
jump its rotation periodically, because of no reason to show that
the critical value $\sigma_{\rm c}$ (and maybe $\mu$) keeps a
constant during its life. The time separation $t_{\rm q}$ may
present a quasi-periodic nature if $\sigma_{\rm c}$ has a Gaussian
distribution.
\begin{figure}[t]
  \centering
  \begin{minipage}[t]{.5\textwidth}
    \centering
  \end{minipage}%
     \caption{
Contour lines of the time separation between two quakes, $t_{\rm
q}$, on the $P-\dot{P}$ diagram. This glitching interval time
$t_{\rm q}$ is labelled to the lines in unit of years. The
parameters are taken as: $M=1.4M_\odot$, $\rho=4\times
10^{14}$g/cm$^3$, and $\mu=10^{32}$ erg/cm$^3$. Results with three
values of $\sigma_{\rm c}$ ($10^{18}$, $10^{21}$, and $10^{24}$
erg/cm$^3$) are for indications. The pulsar data are downloaded
from http://www.atnf.csiro.au/research/pulsar/psrcat.%
}
\end{figure}

\section{The post-starquake behaviors}

When $i-$th quake occurs in a pulsar, the oblateness decreases to
$\varepsilon_{-i}$ at first, while the reference point is changed
to $\varepsilon_i$ from $\varepsilon_{i-1}$. Soon after the
rotational jump ($\varepsilon_{+i}\rightarrow \varepsilon_{-i}$),
the star has a tendency to reach the equilibrium oblateness
$\varepsilon_i$. Certainly, this recovery process depends on the
difference, $\varepsilon_{-i}-\varepsilon_i$, which is shown in
Fig.4 as a function of $E_{\rm therm}$. Increasing oblateness can
obviously contribute an additional spindown role besides that of
magnetodipole radiation, which results in larger $\dot P$.
Such a post-starquake could be observed as a recover behavior
during post-glitch.
From Fig.4, one can see the values of
$\varepsilon_{-i}-\varepsilon_i$ are generally not sensitive to
both $E_{\rm therm}$ and $\sigma_{\rm c}$.
\begin{figure}[t]
  \centering
  \begin{minipage}[t]{.5\textwidth}
    \centering
    \includegraphics[width=9cm]{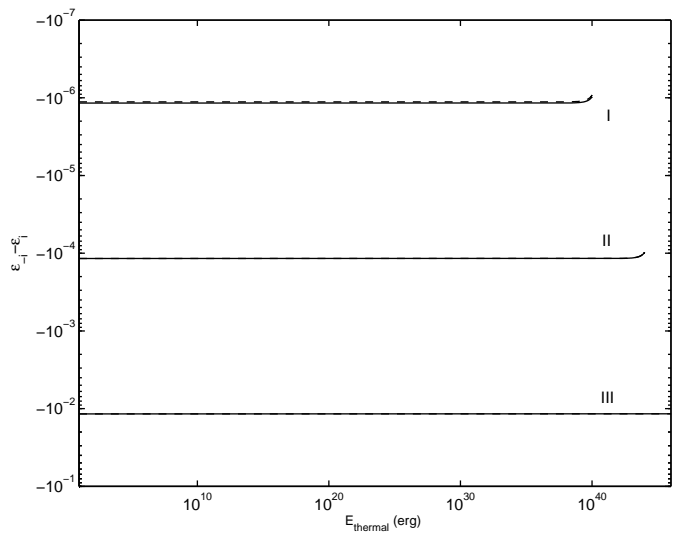}
  \end{minipage}%
     \caption{
The difference of $\varepsilon_{-i}-\varepsilon_{i}$ as a function
of $E_{{\rm therm}}$. The lines are grouped for different periods;
``I'': $P=1$ s, ``II'': $P=0.1$ s, and ``III'': $P=0.01$ s. The
solid and dashed lines are for $\sigma_{\rm c}=10^{16}$ and
$\sigma_{\rm c}=10^{24}$ erg/${\rm cm^{3}}$, respectively, which
are almost the same for certain period $P$. Other parameters
chosen are: $M=1.4M_\odot$, $\rho=4\times 10^{14}$ g/cm$^3$, and
$\mu=10^{32}$ erg/${\rm cm^{3}}$.
}
\end{figure}

The recovery of $\varepsilon_{-i}\rightarrow \varepsilon_i$ is
actually a complex process, in which both elastic revert and
plastic flow could not be negligible. Nevertheless, this recovery
process might be an analog of damped oscillations, with a
mathematical description of
\begin{equation}
\ddot x + 2\beta \dot x + \omega _0 ^2 x = 0,
\label{ddotepsinon}
\end{equation}
with $x\equiv \varepsilon-\varepsilon_i$. The second term in
Eq.(\ref{ddotepsinon}) arise from the anelastic effect during the
recovery. The solution of Eq.(\ref{ddotepsinon}) depends on the
relative strength of the elastic spring and that of the damping.
It is well known that three possible solutions exist; the
so-called underdamped case: $\beta<\omega_0$, the critical case:
$\beta=\omega_0$, and the overdamped case: $\beta>\omega_0$; where
$\omega_0$ is the intrinsic frequency. One comes then to
$\omega_{0}\sim \sqrt{\mu R/M}$ by dimensional analysis (force
$\sim \sigma R^2$ and displacement $\sim x R$).

In all the three cases, the oscillation damps by a factor of $\sim
\exp(-\beta t)$, with a typical time of $\tau=1/\beta$. We are
just to discuss the critical damped motion below for an
indication.
In this case, $\omega_0=\beta$, the time evolution of
$\varepsilon$ after the occurrence of starquake can be solved by
Eq.(\ref{ddotepsinon}),
\begin{equation}
\varepsilon(t) = \varepsilon_i+(\varepsilon_{-i}-\varepsilon
_i)(1+t/\tau)e^{-t/\tau}.
\end{equation}

\section{Conclusions and discussions}

A starquake model for pulsar glitches is developed in the regime
of solid quark stars, and it is found that the general glitch
behaviors (i.e., the glitch amplitude $\Delta\Omega/\Omega$ and
the time interval $t_{\rm q}$) could be reproduced if solid quark
matter has properties of shear modulus $\mu=10^{30\sim 34}$
erg/cm$^3$ and critical stress $\sigma_{\rm c}=10^{18\sim 24}$
erg/cm$^3$.
It is suggested that the post-glitch process could be described as
damped oscillations, especially in the critical and the overdamped
cases.
Anyway, this is only a primary and simplified study of quakes in
solid quark stars, more elaborate work, with possible
modifications, on both quake and postquake processes is necessary
in order to understand the nature of solid quark matter through
glitching pulsars.

We are dealing with solid quark stars in this paper.
The quark Cooper pairing of the BCS type is suggested in quark
matter of low-temperature but high baryon density, which may
result in a color superconducting state \cite{arw98}, with a large
pairing gap on the order of 100 MeV.
This kind of condensation in momentum space takes place in case of
same Fermi momenta; whereas ``LOFF''-like state may occur if the
Fermi momenta of two (or more) species of quarks are different
\cite{abr02}.
For 3 flavors of massless quarks, all nine quarks pair in a
pattern which locks color and flavor symmetries, as called
color-flavor locking (CFL) state \cite{arw99}.
However, for such quark matter, there exists a {\em competition}
between color superconductivity and solidification, just like the
case of laboratory low-temperature physics.
One needs weak-interaction and low-mass in order to obtain a
quantum fluid before solidification. This is why only helium, of
all the elements, shows superfluid phenomenon though other noble
elements have similar weak strength of interaction due to filled
crusts of electrons.
The strong color interaction (and the Coulomb interaction in the
system with strangeness) may be responsible for a possible
solidification of dense quark matter with low temperatures.
Further experiments (in low-energy heavy ion colliders) may answer
whether quark matter is in a state of solid or
color-superconductivity.
Can a solid neutron star be possible? The answer might be {\em
no}, because at least the part of neutron matter with approximate
nuclear saturation density should be in a fluid state.
In this sense, only solid quark matter is possible, and a quark
star is identified if one convinces that a pulsar is in a solid
state.

We have assumed that the entire strain energy $E_{\rm strain}$ is
relieved in a quake, which results in the reference oblateness of
Eq.(\ref{epsiloni}).
However, it is possible that not entire, but most of, the energy
$E_{\rm strain}$ is released in a real situation, and the actual
reference points are near but larger than $\varepsilon_i$. Of
course, there is a tendency of $\varepsilon\rightarrow
\varepsilon_i$ after the $i-$th quake, but the effective shear
modulus, $\mu_{\rm eff}$, of matter broken could be much smaller
than that of perfect elastic solid, $\mu$.
The recovery timescale could be $\tau \sim 15$ days if $\mu_{\rm
eff}$ is order of $10^{15}$ erg/cm$^3$.

A large Vela glitch on 2000 January 16.319 was noted
\citep{dmc00}, and {\em Chandra} observations were carried out
$\sim 3.5$ and $\sim 35$ days after the glitch \citep{hgh01}, but
no temperature change expected in conventional models with
released thermal energy of $\sim 10^{42}$ ergs is detected.
This could be understood in this starquake model since (1) the
thermal conductivity of quark matter is much larger than that of
hadron matter and (2) the thermal energy released, $E_{\rm
therm}$, to be much smaller than $10^{42}$ ergs is possible.

{\em Acknowledgments}:
This work is supported by National Nature Sciences Foundation of
China (10273001), and by the Special Funds for Major State Basic
Research Projects of China (G2000077602). The valuable suggestions
from an anonymous referee are sincerely acknowledged.

\end{document}